\documentclass[letterpaper, 10 pt, conference]{ieeeconf}  

\usepackage{amsmath}
\usepackage{amssymb} 

\renewcommand{\QED}{\QEDopen}

\usepackage[labelformat=simple]{subcaption}
\DeclareCaptionLabelSeparator{periodspace}{.\quad}
\usepackage{tikz}
\usetikzlibrary{arrows,matrix,positioning,patterns}
\usetikzlibrary{calc}
\usepackage{pgfplots} 
\usetikzlibrary{plotmarks}

\usepackage[utf8]{inputenc}
\usepackage{pgfplots}
\usepgfplotslibrary{groupplots}

\usepackage{url}

\usepackage{multirow}
\usepackage{colortbl}

\usepackage[ruled, vlined, linesnumbered]{algorithm2e}

\SetCommentSty{mycommfont}

\DeclareSymbolFont{bbold}{U}{bbold}{m}{n}
\DeclareSymbolFontAlphabet{\mathbbold}{bbold}

\usepackage[labelformat=simple]{subcaption}
\DeclareCaptionLabelSeparator{periodspace}{.\quad}
\usepackage{listings}
\usepackage{xcolor}
\usepackage{cite}

\usepackage{wrapfig}
\usepackage{supertabular}

\IEEEoverridecommandlockouts                              
\title{\LARGE \bf
Integrated Time Series Summarization and Prediction Algorithm\\ and its Application to COVID-19 Data Mining}

\author{Mogens Graf Plessen
\thanks{{\tt\small mgplessen@gmail.com}}}
\renewcommand{\baselinestretch}{0.88} 
\begin{document}

\maketitle
\thispagestyle{empty}
\pagestyle{empty}

\begin{abstract}
This paper proposes a simple method to extract from a set of multiple related time series a compressed representation for each time series based on statistics for the entire set of all time series. This is achieved by a hierarchical algorithm that first generates an alphabet of shapelets based on the segmentation of centroids for clustered data, before labels of these shapelets are assigned to the segmentation of each single time series via nearest neighbor search using unconstrained dynamic time warping as distance measure to deal with non-uniform time series lenghts. Thereby, a sequence of labels is assigned for each time series. Completion of the last label sequence permits prediction of individual time series. Proposed method is evaluated on two global COVID-19 datasets, first, for the number of daily net cases (daily new infections minus daily recoveries), and, second, for the number of daily deaths attributed to COVID-19 as of April 27, 2020. The first dataset involves 249 time series for different countries, each of length 96. The second dataset involves 264 time series, each of length 96. Based on detected anomalies in available data a decentralized exit strategy from lockdowns is advocated. 
\end{abstract}

\section{Introduction\label{sec_intro}}

Data mining is the task to extract higher level information from raw data \cite{hand2007principles}. In general, data can be available in manifold form. In this paper, the focus is on \emph{time series} data mining \cite{fu2011review}.

Given a set of multiple time series it is desirable in many applications to extract information in compressed or simplified form while still being representative. This task can be referred to as \emph{summarization}, where essential features are retained that ``fit on a single page, computer screen, or executive summary, etc'' \cite{lin2003symbolic}. Another popular task is time series \emph{prediction} \cite{weigend2018time}. It is natural to ask for a method that addresses both combinedly. There are two basic approaches:
\emph{model-based} and \emph{model-free} algorithms. For the former, ``model fitting'' or ``system identification'' is the term used in the automatic control field for estimating \emph{dynamical models} based on measurements of the system's input and output signals \cite{ljung1999system}, which can then be used for predction and control. The disadvantage is that a priori assumptions (an ``educated guess'') on the mathematical model need first to be made, which requires data domain expertise. Therefore, the second model-free approach is here preferred.

Henceforth, the motivation and contribution of this paper is to present an integrated algorithm for both model-free time series summarization and prediction. The second aspect is its application to real world data from the recent COronaVIrus Disease 2019 (COVID-19) for both global net daily infections and daily deaths for the time period from January 23 until April 27, 2020.

\begin{table}
\vspace{0.3cm}
\centering
\begin{tabular}{|ll|}
\hline
\multicolumn{2}{|c|}{MAIN NOMENCLATURE}\\
\multicolumn{2}{|l|}{Symbols}\\

$\bar{c}_k(t)$ & Centroid time series for cluster $k$ at time $t$.\\
$l_i(p)$ & Label assigned to time series $i$ for segment $p$.\\
$l_i^\text{pred}$ & Predicted label for the next segment of time series $i$.\\
$\Delta t_i$ & Prediction horizon for time series $i$.\\
$\Delta \tau^\text{min}$ & Hyperparameter, minimum segment length.\\
$\tau_i(p)$ & Segmentation time for time series $i$ and segment $p$.\\
$\bar{\tau}_k(q)$ & Segmentation time for centroid $k$ and segment $q$.\\
$x_i(t)$ & Data point for time series $i$ at time $t$.\\
$x_i^\text{pred}(t)$ & Prediction for time series $i$ at time $t\geq T$.\\
$\mathcal{I}(i)$ & Indexing assignment of time series $i$ to a cluster.\\
$K^\text{max}$ & Hyperparameter, maximum number of clusters.\\
$K$ & Number of clusters obtained after $K$-means\text{++}.\\
$N$ & Number of different time series.\\
$S^\text{max}$ & Hyperparameter, maximum number of segments.\\
$T$ & Length of time series.\\[3pt]
\multicolumn{2}{|l|}{Abbreviations}\\
ISPA & Integrated Summarization and Prediction Algorithm.\\
$K$-means\text{++} & Patritioning clustering algorithm \cite{arthur2006k}. \\
APTS & A Posteriori Trading-inspired Segmentation \cite{plessen2019posteriori}. \\
DTW & Dynamic Time Warping \cite{sakoe1978dynamic}. \\
\hline
\end{tabular}
\end{table}

For many different model-free time series data mining tasks there are three  essential \emph{building blocks} (basic analysis techniques) that frequently recur throughout the literature, and which will also serve as foundation for proposed method. These are \emph{clustering}, \emph{segmentation} and \emph{similarity measures}. Related literature is here briefly summarized. 

The well-known \emph{$K$-means} algorithm \cite{lloyd1982least} does not offer any accuracy guarantess, but its simplicity and speed make it appealing for clustering. Furthermore, by modification of the initialization scheme the  algorithm called \emph{$K$-means}\texttt{++} can be obtained which is $\mathcal{O}(\log K)$-competitive with the optimal clustering \cite{arthur2006k}. Typically, the number of clusters $K$ is prescribed as a hyperparameter. However, there are methods to also learn $K$, for example, via the ``Elbow method'' or the ``Silhouette method'' \cite{rousseeuw1987silhouettes}. Variants of the $K$-means algorithm belong to the class of partitioning clustering. In general, there are other clustering classes such as hierarchical, spectral, density-based clustering and more \cite{berkhin2006survey}. For a survey on clustering for time series see \cite{liao2005clustering}. 

Time series segmentation can be useful for a plethora of applications such as dimensionality reduction for the handling of very long time series or motif discovery. It is also closely related to (and may in the literature be interchangeably called) ``changepoint detection'', ``breakpoint detection'' or ``event detection'' \cite{aminikhanghahi2017survey}. Segment boundaries can be interpreted as the changepoints where characteristic behavior changes. Typically, the number of segments or a desired upper bound thereof is specified as a hyperparameter. It can be further distinguished between model-based and model-free approaches. For example, for the former class, in \cite{liu2008novel} piecewise affine time series approximations are used for their segmentation logic. A recent model-free approach inspired by a posteriori optimal trading was presented in \cite{plessen2019posteriori}.

The choice of similarity measure (also referred to as distance measure) \cite{cassisi2012similarity} plays an important role in many time series data mining algorithms, including for subsequence matching  and similarity search \cite{faloutsos1994fast}, time series classification \cite{bagnall2017great} and also for aforementioned clustering. By far the most common similarity measure used for time series is the Euclidean distance (ED). Another popular choice is \emph{dynamic time warping} (DTW) \cite{sakoe1978dynamic}. It has shown excellent performance on supervised time series classification tasks in combination with ``nearest neighbor search'' when its hyperparameter--the \emph{warping window}--is learned on a training dataset (typically via ``leave-one-out cross-validation'') \cite{tan2018efficient}. Furthermore and in contrast to ED, DTW offers the advantage to compute a distance between two time series of different lengths. 

The remaining paper is organized as follows. The problem is formulated mathematically in \S \ref{sec_problFormulation}. The proposed solution is presented in \S \ref{sec_problem_soln}. Numerical results are evaluated in \S \ref{sec_expts}, before concluding with \S \ref{sec_concl}.

\section{Problem Formulation\label{sec_problFormulation}}

The problem addressed is twofold; First, the development of an integrated data mining algorithm for summarization and prediction that extracts information from a set of multiple related time series. Second, its application to real-world data from  the COronaVIrus Disease 2019 (COVID-19) \cite{covid19data}. While time series differ for different countries, all COVID-19 time series are related by a common cause. Two case studies for both global net daily infections and daily deaths serve for evaluation of proposed method. 

Given a set of $N$ time series of uniform length $T$, $\{\{x_i(t)\}_{t=0}^{T-1}\}_{i=0}^{N-1}$, with $x_i(t)\in\mathbb{R}$, an algorithm is sought that returns four primary outputs. These are a compressed representation of each time series that consists of, first, a sequence of labels, $\{\{l_i(p)\}_{p=0}^{P_i-1}\}_{i=0}^{N-1}$, with $l_i(p)\in\mathbb{N}_+$, second, a sequence of segmentation time instants, $\{\{\tau_i(p)\}_{p=0}^{P_i}\}_{i=0}^{N-1}$, with $\tau_i(p)\in\mathbb{N}_+$, $\tau_i(0)=0$ and $\tau_i(P_i)=T-1$, third, a prediction for each time series, $\{\{x_i^\text{pred}(t)\}_{t=T-1}^{T-1+\Delta t_i}\}_{i=0}^{N-1}$ with prediction horizon $\Delta t_i\in\mathbb{N}_{++}$, which in general may vary for each time series $i=0,\dots,N-1$, and, finally, corresponding prediction labels $\{ l_i^\text{pred}\}_{i=0}^{N-1}$. Secondary outputs, which can be considered as a byproduct of the main algorithm, are, first, a set of $K$ centroid time series, $\{\{\bar{c}_k(t)\}_{t=0}^{T-1}\}_{k=0}^{K-1}$, with $\bar{c}_k(t)\in\mathbb{R}$, second, a sequence of segmentation time instants for these centroids, $\{\{\bar{\tau}_k(q)\}_{q=0}^{Q_k}\}_{k=0}^{K-1}$, and third, assignments of each time series to clusters associated with each centroid, $\mathcal{I}(i)\in\{0,\dots,K-1\},~\forall i=0,\dots,N-1$.

\section{Problem Solution\label{sec_problem_soln}}

This section is organized by first stating proposed algorithm, before elaborating on the design of substeps.

\subsection{Integrated Summarization and Prediction Algorithm\label{subsec_ProposedSoln}}

Algorithm \ref{alg_ISPA} summarizes suggested method and is called  ISPA (Integrated Summarization and Prediction Algorithm).

For time series mining of raw data the common preprocessing step is to \emph{z-normalize}. Thus, denoting raw data with a superscript, data input to Algorithm \ref{alg_ISPA} typically is $x_i(t)=\frac{x_i^\text{raw}(t)-\mu_i}{\sigma_i},\forall t,\forall i,$ with $\mu_i=\frac{1}{T}\sum_{t=0}^{T-1}x_i^\text{raw}(t)$ and $\sigma_i^2=\frac{1}{T}\sum_{t=0}^{T-1}x_i^\text{raw}(t)^2 - \mu_i^2$. However, for COVID-19 data it was observed that \emph{omission} of z-normalization clearly clarified clustering results. This is explained by the fact that available data is so strongly discriminating for different countries with variations in orders of magnitude. Some regions report numbers in single digits, while other countries report in the hundreds or even thousands.

The objective of Steps 1-3 in Algorithm \ref{alg_ISPA} is to learn an \emph{alphabet} of overall data-characterizing \emph{shapelets} (short segments or subsequences). These are obtained from centroids after a clustering step. Therefore, data input, $\{\{x_i(t)\}_{t=0}^{T-1}\}_{i=0}^{N-1}$, is clustered via aforementioned $K$-means\texttt{++}  algorithm \cite{arthur2006k}. This step maps the data input to a set of $K$ centroids, $\{\{\bar{c}_k(t)\}_{t=0}^{T-1}\}_{k=0}^{K-1}$, whereby the \emph{resulting} number of obtained clusters can be less than a provided hyperparameter upper bound, i.e., $K\leq K^\text{max}$. This is since an unconstrained $K$-means\texttt{++} implementation permits in an overall cost-minimizing fashion to return some empty-valued clusters (which are consequently discarded), in case $K^\text{max}$ is selected conservatively large or the underlying data is very discriminating (which is the case for COVID-19). Step 1 of Algorithm \ref{alg_ISPA} also returns $\{\mathcal{I}(i)\}_{i=0}^{N-1}$, the indexing assignment of each time series to a cluster with $\mathcal{I}(i)\in\{0,\dots,K-1\}$. While this information is not exploited further, it is kept as part of the time series summarization. 

\begin{algorithm}
\SetKwInOut{Subfunctions}{\textbf{Subfunctions}}
\SetKwInOut{Input}{\textbf{Data Input}}
\SetKwInOut{Hyperparameters}{\textbf{Hyperparam.\hspace{0.05cm}}}
\SetKwInOut{POutput}{\textbf{Main Output}}
\SetKwInOut{AuxOutput}{\textbf{Other}}
\DontPrintSemicolon
\vspace{0.15cm}
\Subfunctions{$\mathcal{F}^\text{$K$-means++}(\cdot)$, $\mathcal{F}^\text{APTS}(\cdot)$, $\mathcal{F}^\text{DTW}(\cdot)$.}
\vspace{0.15cm}\hrule\vspace{0.15cm}
\Hyperparameters{$K^\text{max}$, $S^\text{max}$, $\Delta \tau^\text{min}$.}
\vspace{0.15cm}\hrule\vspace{0.15cm}
\Input{$\{\{x_i(t)\}_{t=0}^{T-1}\}_{i=0}^{N-1}$.}
\vspace{0.15cm}\hrule\vspace{0.15cm}
\tcp{\textbf{A. CREATE ALPHABET OF LABELS \& SHAPELETS.}}\vspace{0.15cm}
$K,\{\{\bar{c}_k(t)\}_{t=0}^{T-1}\}_{k=0}^{K-1},\{\mathcal{I}(i)\}_{i=0}^{N-1} \leftarrow ~\dots\newline  \hspace*{1.5cm}\dots~\mathcal{F}^\text{$K$-means++}( \{\{x_i(t)\}_{t=0}^{T-1}\}_{i=0}^{N-1},K^\text{max}).$\;
\For{$k\in\{0,\dots,K-1\}$}
{
$\{\bar{\tau}_k(q)\}_{q=0}^{Q_k} \leftarrow \mathcal{F}^\text{APTS}( \{\bar{c}_k(t)\}_{t=0}^{T-1},S^\text{max},\Delta \tau^\text{min} ) .$\;
} 
\vspace{0.15cm}\hrule\vspace{0.15cm} 
\tcp{\textbf{B. ASSIGN LABEL SEQUENCE TO ALL CHANNELS.}}\vspace{0.15cm}
\For{$i\in\{0,\dots,N-1\}$}
{
$\{\tau_i(p)\}_{p=0}^{P_i} \leftarrow \mathcal{F}^\text{APTS}( \{x_i(t)\}_{t=0}^{T-1},S^\text{max},\Delta \tau^\text{min} ).$\;
\For{$p\in\{0,\dots,P_i-1\}$}
{
$k^\star,q^\star \leftarrow \text{arg}\min\limits_{\substack{0\leq q < Q_k\\0\leq k < K}} \mathcal{F}^{DTW}\big( \dots\newline \hspace*{1.2cm} \dots ~\{x_i(t)\}_{t=\tau_i(p)}^{\tau_i(p+1)}, \{\bar{c}_k(t)\}_{t=\bar{\tau}_k(q)}^{\bar{\tau}_k(q+1)} \big).$\;
$ l_i(p) \leftarrow k^\star S^\text{max}+q^\star.$\;
}
} 
\vspace{0.15cm}\hrule\vspace{0.15cm} 
\tcp{\textbf{C. SEGMENT PREDICTION FOR EACH CHANNEL.}}\vspace{0.15cm}
\For{$i\in\{0,\dots,N-1\}$}
{
$k\leftarrow \lfloor \frac{l_i(P_i-1)}{S^\text{max}} \rfloor$.\;
$q\leftarrow l_i(P_i-1)-k S^\text{max}$.\;
$l_i^{\text{pred}} \leftarrow l_i(P_i-1) + 1$.\;
$q\leftarrow q+1$.\;
$\Delta t_i\leftarrow \bar{\tau}_k(q+1)-\bar{\tau}_k(q)$.\;
$\{x_i^\text{pred}(t)\}_{t=T-1}^{T-1+\Delta t_i} \leftarrow ~ \dots \newline ~~~~~~\dots~\{ x_i(T-1) - \bar{c}_k(\bar{\tau}_k(q)) + \bar{c}_k(t)  \}_{t=\bar{\tau}_k(q)}^{\bar{\tau}_k(q+1)} $.\;
}
\vspace{0.15cm}\hrule\vspace{0.15cm}
\POutput{$ \{ \{l_i(p)\}_{p=0}^{P_i-1} \}_{i=0}^{N-1},~\{\{\tau_i(p)\}_{p=0}^{P_i}\}_{i=0}^{N-1},\newline \{\{x_i^\text{pred}(t)\}_{t=T-1}^{T-1+\Delta t_i}\}_{i=0}^{N-1},~\{l_i^{\text{pred}}\}_{i=0}^{N-1}$.}
\AuxOutput{$ K,~\{\{\bar{c}_k(t)\}_{t=0}^{T-1}\}_{k=0}^{K-1},~\{ \bar{\tau}_k(q)\}_{q=0}^{Q_k},\newline  \{\mathcal{I}(i)\}_{i=0}^{N-1}$.}
%
%
\caption{ISPA}\label{alg_ISPA}
\end{algorithm}
\vspace{-0.2cm}

In Steps 3 the alphabet of shapelets is generated by segmenting all centroids with APTS \cite{plessen2019posteriori}. Here, a simplified basic implementation with a reduced number of hyperparameters is employed, which is summarized in Algorithm \ref{alg_APTSbasic}. In Step 1 a normalization is carried out in order to render time series data positive, which is a prerequisite for used segmentation logic. First, channel-wise data is z-normalized, before a linear transformation is applied, $\tilde{x}_i(t)=x_i(t) + \tilde{x}_i^\text{offset},\forall t=0,\dots,T-1$, with $\tilde{x}_i^\text{offset}=|\text{min}_{t\in\{0,\dots,T-1\}} x_i(t)| + 1$.  

The fundamental idea of APTS is to treat normalized time series as a surrogate stock that can be traded optimally a posteriori in virtual portfolio holding either cash or stock \cite{plessen2018posteriori}. The resulting rebalancing time instances are recorded as the segmentation time instants, which consequently typically appear at local peaks and bottoms. For ease of reference the derivation of the main mapping, $\mathcal{F}^\text{trade}(\cdot)$ in Algorithm \ref{alg_APTSbasic}, is here repeated from \cite[\S III.A]{plessen2019posteriori}. 

Channel-wise surrogate wealth dynamics are introduced that model a virtual portfolio holding either a virtual cash position or the ``stock'' modeled by the channel-wise normalized time series. Therefore, a four-dimensional state vector is defined, $z_i(t)=[n_i(t),~c_i(t),~b_i(t),~w_i(t)]$, where $n_i(t)\geq 0$  models the number of shares held at time $t$, $c_i(t)\geq 0$ the cash position, $b_i(t)\in\{-1,1\}$ the binary state indicating full investment in cash or stock, respectively, and finally $w_i(t)\geq 0$ denoting total wealth at time $t$. In contrast to regular stock trading for our surrogate setup the number of shares, $n_i(t)\geq 0$, is real-valued. At $t=0$ the state vector is initialized with $z_i(0)=[0,~\frac{\tilde{x}_i(0)}{1-\epsilon_i},~-1,~\frac{\tilde{x}_i(0)}{1-\epsilon_i}]$, where $\epsilon_i\in[0,1)$ is interpreted as a linear transaction cost level. This initialization implies an initial cash position sufficient to buy one share when accounting for transaction cost and is defined this way to avoid adding scaling hyperparameters. Then, introducing a control variable, $u_i(t)\in\{-1,1\}$, state transition dynamics are defined as,
\begin{equation}
z_i(t+1) = \begin{cases} 
\begin{bmatrix} 0\\ c_i(t)\\-1\\c_i(t)\end{bmatrix}, & ~\text{if}~u_i(t)=-1,\\
\begin{bmatrix}\frac{c_i(t)(1-\epsilon_i)}{\tilde{x}_i(t)}\\0\\1\\n_i(t+1)\tilde{x}_i(t+1)\end{bmatrix}, & ~\text{if}~u_i(t)=1,
\end{cases}
\label{eq_def_zitp1_a}
\end{equation}
for $z_i(t)\in\{z_i(t)\in\mathbb{R}^4:b_i(t)=-1\}$, and
\begin{equation}
z_i(t+1) = \begin{cases} 
\begin{bmatrix}0\\n_i(t)\tilde{x}_i(t)(1-\epsilon_i)\\-1\\c_i(t+1)\end{bmatrix}, & ~\text{if}~u_i(t)=-1,\\
\begin{bmatrix}n_i(t)\\0\\1\\n_i(t)\tilde{x}_i(t+1)\end{bmatrix}, & ~\text{if}~u_i(t)=1,
\end{cases}
\label{eq_def_zitp1_b}
\end{equation}
for $z_i(t)\in\{z_i(t)\in\mathbb{R}^4:b_i(t)=1\}$. Combinedly, \eqref{eq_def_zitp1_a}-\eqref{eq_def_zitp1_b} model all four possible transitions between full cash and full stock investment subject to linear transaction costs. In a causal setting $\tilde{x}_i(t+1)$ is not known at time $t$. However, in the batch setting it is available, which is equivalent to perfect one step-ahead knowledge a posteriori. Therefore, and due to the fact of a positive $\tilde{x}_i(t)$ by above discussion there always exists a wealth-maximizing trading trajectory from $t=0$ to $T$ as a function of transaction cost level $\epsilon_i\geq 0$ such that $w_i(T)$ is channel-wise maximized. This trajectory can be computed efficiently as follows. Starting from $z_i(0)$ defined above, at $t=1$ two possible states can result which differ by binary $b_i(1)\in\{-1,1\}$. Let these two states be denoted by $z_i^{(-1)}(1)$ and $z_i^{(1)}(1)$, respectively. Then, by Bellman's principle of optimality and with the purpose of deriving the wealth-maximizing trading trajectory the following recursion can be implemented for all $t\geq 1$. When $\tilde{x}_i(t+1)$ becomes available, $z_i^{(-1)}(t)$ and $z_i^{(1)}(t)$ branch out to a total of four different states according to \eqref{eq_def_zitp1_a}-\eqref{eq_def_zitp1_b}. These are pruned to two by selecting the $w_i(t+1)$-maximizing solutions for each $b_i(t+1)=-1$ and $b_i(t+1)=1$ such that $z_i^{(-1)}(t+1)$ and $z_i^{(1)}(t+1)$ are obtained, respectively. Their corresponding optimal parent states are further recorded, which shall be denoted by $z_i^{(-1),\text{parent}}(t)$ and $z_i^{(1),\text{parent}}(t)$. This recursion is repeated until $t=T$. Given $z_i^{(-1)}(T)$ and $z_i^{(1)}(T)$ let $b_i^\star(T)=-1$ if $w_i^{(-1)}(T)> w_i^{(1)}(T)$, and $b_i^\star(T)=1$ otherwise. Now, using the list of optimal parent states, $\{z_i^{(-1),\text{parent}}(t)\}_{t=1}^{T-1}$ and $\{z_i^{(1),\text{parent}}(t)\}_{t=1}^{T-1}$, the optimal wealth-maximizing trading trajectory can be obtained by backpropagation, resulting in $\{b_i^\star(t)\}_{t=0}^T$. In the following, $\mathcal{F}^\text{trade}(\{ \tilde{x}_i(t) \}_{t=0}^T,\epsilon_i)$ shall abbreviate the mapping from $\{ \tilde{x}_i(t) \}_{t=0}^T$ to $\{b_i^\star(t)\}_{t=0}^T$ as a function of $\epsilon_i>0$. 

Via the choice of $\epsilon_i$ the frequency of  segmentation time instants can be controlled. The larger $\epsilon_i$ the fewer segments. This explains the design of Steps 2-12 in Algorithm \ref{alg_APTSbasic} in combination with user-provided hyperparameter $S^\text{max}$, which defines the maximum desired number of segmentation time instants. Starting from $\epsilon_i=0$ the corresponding number of segments is computed. Should this exceed $S^\text{max}$ the transaction cost proxy $\epsilon_i$ is increased according to Step 12. By default the step size is here selected as $\Delta \epsilon_i=0.01$. The iteration in Algorithm \ref{alg_APTSbasic} is guaranteed to terminate for $S^\text{max}>2$ since the number of segments is guaranteed to be monotonously decreasing with increasing $\epsilon_i$ \cite{plessen2019posteriori}.  

Hyperparameter $\Delta \tau^\text{min}$ in Step 5 and 9 was introduced specifically for volatile COVID-19 data to ensure that the last segment (which is consequently used for prediction via nearest neighbor search as outlined below) has a minimum length of at least $\Delta \tau^\text{min}$ time steps.

Shapelets returned by Steps 1-3 of ISPA are given by $\{\bar{c}_k(t)\}_{t=\bar{\tau}_k(q)}^{\bar{\tau}_k(q+1)},\forall q=0,\dots,Q_k-1,\forall k=0,\dots,K-1$, wereby boundary segmentation time instants are defined with $\bar{\tau}_k(0)=0$ and $\bar{\tau}_k(Q_k)=T-1$ for all $k=0,\dots,K-1$. Shapelets are indexed with $k S^\text{max}+q,\forall k=0,\dots,K,\forall q=0,\dots,S^\text{max}$. Eventhough it typically results $Q_k < S^\text{max}$, the indexing method using $S^\text{max}$ is used in order to retain a simple formula (instead of defining an additional array for storage of the number of segments for each cluster). 

Steps 4-8 of Algorithm \ref{alg_ISPA} determine the segmentations for each of the $N$ time series, before the closest shapelet from the alphabet of shapelets is assigned to each segment via a nearest neighbor search for a given similarity measure. In detail, after application of APTS, segments are described by $\{ x_i(t) \}_{t=\tau_i(p)}^{\tau_i(p+1)},\forall  p=0,\dots,P_i-1,\forall i=0,\dots,N-1$. In order to determine the closest shapelet from the set of all centroid segments (the alphabet of reference segments), a standard \emph{nearest neighbor search} is conducted employing unconstrained DTW to compute a quadratic error function for time series of different lengths. Because of its widespread usage DTW is here not further discussed, and it is referred to \cite{sakoe1978dynamic}. Two comments are made. First, \emph{unconstrained} DTW is suitable since segments typically include very few time instants. This is beneficial in the sense that the otherwise typical hyperparameter, the \emph{warping window}, does not have to be learned (in a training phase) or set artificially, which otherwise can be computationally expensive to do \cite{tan2018efficient}. Second, Euclidean distance (ED) is not suitable as similarity measure. This is because of its \emph{metric}-nature \cite{cassisi2012similarity}. ED does not permit to compare two time series of different lengths. 

Once the closest shapelet is determined and indexed by $k^\star$ and $q^\star$, label $l_i(p)$ is defined as in Step 8 of Algorithm \ref{alg_ISPA}. Then, the collection of multiple of these labels, $\{l_i(p)\}_{p=0}^{P_i}$, defines a \emph{word} and defines a compressed time series representation for time series $i=0,\dots,N-1$. Note that even if time series $i$ is assigned to cluster $k$, i.e., $\mathcal{I}(i)=k$, this does not necessarily imply that the label transitions for $\{l_i(p)\}_{p=0}^{P_i}$ must coincide with the label transitions for the segments of the corresponding centroid. 

\begin{algorithm}
\SetKwInOut{Subfunctions}{\textbf{Subfunctions}}
\SetKwInOut{Input}{\textbf{Data Input}}
\SetKwInOut{Hyperparameters}{\textbf{Hyperparam.\hspace{0.05cm}}}
\SetKwInOut{Output}{\textbf{Final Result}}
\DontPrintSemicolon
\vspace{0.15cm}
\Subfunctions{$\mathcal{F}^\text{normalize}(\cdot)$ and $\mathcal{F}^\text{trade}(\cdot)$.}
\vspace{0.15cm}\hrule\vspace{0.15cm}
\Hyperparameters{$S^\text{max}$, $\Delta \tau^\text{min}$, default $\Delta \epsilon_i=0.01$.}
\vspace{0.15cm}\hrule\vspace{0.15cm}
\Input{$\{x_i(t)\}_{t=0}^{T-1}$.}
\vspace{0.15cm}\hrule\vspace{0.15cm}
$\{ \tilde{x}_i(t) \}_{t=0}^{T-1} \leftarrow \mathcal{F}^\text{normalize}\left( \{ x_i(t) \}_{t=0}^{T-1} \right)$, $\epsilon_i\leftarrow 0$.\;
\While{continue}
{
$\{ b_i^{\star}(t) \}_{t=0}^{T-1} \leftarrow \mathcal{F}^\text{trade}\left( \{\tilde{x}_i(t)\}_{t=0}^{T-1},\epsilon_i\right)$.\;
$p=1$, $t=1$, $\tau_i(0)=0$.\;
\While{$t<T-\Delta \tau^\text{min} ~\&\& ~p<S^\text{max}-1$}
{
\If{$b_i^\star(t+1)\neq b_i^\star(t)$}
{
$\tau_i(p)=t$, $p=p+1$.\;
}
$t=t+1$.\;
}
\If{$t==T-\Delta \tau^\text{min}$}
{
$\tau_i(p)=T-1$, $P_i=p$ and \emph{break}.\;
}\Else
{
$\epsilon_i=\epsilon_i + \Delta \epsilon_i$.\;
}
} 
\vspace{0.15cm}\hrule\vspace{0.15cm}
\Output{$\{ \tau_i(p) \}_{p=0}^{P_i}$.}
%
%
\caption{APTSbasic (per channel $i$)}\label{alg_APTSbasic}
\end{algorithm}
\vspace{-0.2cm}

While Steps 1-8 of ISPA are concerned to structure and summarize the data input, Steps 9-15 define a method to exploit gained information for \emph{prediction} of all time series from the last available time $T-1$ on. Steps 10-11 map from the label $l_i(P_i-1)$ associated with the last known segment of time series $i$ to the closest reference shapelet defined by cluster index $k$ and centroid segment index $q$. Then, in Step 12  the next predicted label, $l_i^\text{pred}$, is determined as the next corresponding centroid segment. Ultimately, the predicted time series, $\{ x_i^\text{pred}(t) \}_{t=T-1}^{T-1+\Delta t_i}$, is generated by normalizing the appropriate reference shapelet and attaching it to $x_i(T-1)$ as indicated in Step 14-15.

\subsection{Discussion and unsupervised learning perspective\label{subsec_ISPA_discussion}}

The novelty of Algorithm \ref{alg_ISPA} stems from its method of combining three main building blocks. These are the algorithms $K$-means\texttt{++}\cite{arthur2006k}, APTS \cite{plessen2019posteriori} and DTW \cite{sakoe1978dynamic}. The selection of unconstrainted DTW for nearest neighbor search is appropriate since it permits to deal with time series segments of different lengths. The benefits of APTS over alternative segmentation algorithms, in particular, also for computational efficiency, were demonstrated in \cite{plessen2019posteriori}. (The simplified version presented here in Algortihm \ref{alg_APTSbasic} is even faster.) The selection of $K$-means\texttt{++} for clustering is appropriate since it requires only a single hyperparameter and provides various useful summarization information such as centroids and cluster assignments for all time series. Extracting shapelets from centroid segments permits generalizing over all available time series, while simultaneously limiting the size of the alphabet of shapelets generated. The complexity of $K$-means is $\mathcal{O}(IKNT)$, where $I$ denotes the number of iterations needed until convergence, $K$ the number of clusters, $N$ the number of time series, and $T$ the length of each time series \cite{arthur2006k}. The complexity of unconstrained DTW is $\mathcal{O}(T_1 T_2)$ for two time series of lengths $T_1$ and $T_2$ \cite{sakoe1978dynamic}. (Note that for the given setting comparing short time series \emph{segments}, it typically is $T_1,T_2\ll T$.) The complexity of APTS is $\mathcal{O}(N_\epsilon T)$ \cite{plessen2019posteriori}, with $N_\text{epsilon}$ denoting the number of $\epsilon$-iterations required until $P_i<S^\text{max}$ in Algortihm \ref{alg_APTSbasic}, which in practice is very small. The dominating complexity of ISPA is dependent on data dimensions. It is either $\mathcal{O}(IK^\text{max}NT)$ for the clustering Step 1 in Algorithm \ref{alg_ISPA}, or $\mathcal{O}(N(N_\epsilon T + N (S^\text{max})^2 T_{ip} T_{kq} K^\text{max}))$ for the nearest neighbor search in Steps 7-13 with $T_{ip}=\max_{i,p} \tau_i(p+1)-\tau_i$ and $T_{kq}=\max_{k,q} \tau_k(q+1) -\tau_k(q)$. The latter is a worst-case upper bound. Typically $T_{ip},T_{kq}\ll T$ and early abandoning is used to accelerate nearest neighbor search. For final hypereparameter selections and given data the computational time is distributed with 62\% to clustering Step 1, 35\% to Step 4-8, and 3\% for other.   
 
ISPA is an unsupervised learning algorithm with applications to a set of multiple related unlabeled time series. In contrast to supervised settings, such as \emph{time series classification} \cite{bagnall2017great}, there does not exist standard ground-truth data that permits straightforwad evaluation and comparison of algorithms. In the unsupervised setting evaluations of comparative methods are subjective. Either an artificially made-up ranking measure has to be designed or solutions have to be plotted and compared visually according to aesthetic criteria. The former can be misleading via biased measure design towards different objectives of algorithms to be compared. For visual comparison of prediction methods a \emph{Martingale process} is considered, using the present value as future mean estimate. Computational efficiency is demonstrated by fast overall solve times (less than 1s) on a 2014-old laptop. Finally, all main three subroutines ($K$-means\texttt{++}, APTS and DTW) offer guarantees about termination in finite time \cite{sakoe1978dynamic,plessen2019posteriori,arthur2006k}. Therefore, ISPA also offers this guarantee. All methods were implemented in C++. All experiments were run on an Intel i7-4810MQ CPU@2.8GHz$\times$8 processor with 15.6 GiB memory. Runtimes of Algorithm \ref{alg_ISPA} for the two experiments discussed below were 0.6s and 0.89s, respectively.

\begin{table}
\renewcommand{\arraystretch}{0.9}
\vspace{0.6cm}
\centering
\begin{tabular}{|l|l|l|}
\hline
 Symbol & Value\\ 
\hline
$K^\text{max}$ & N \\ 
$S^\text{max}$ & 10 \\  
$\Delta \tau^\text{min}$ & 5\\  
\hline
\end{tabular}
\caption{Selection of hyperparameters. To demonstrate the naturally discriminating characteristic of COVID-19 data it was selected $K^\text{max}=N$. Numerical values are 249 and 264 for Experiment 1 and 2, respectively. }
\label{tab_hyperparam}
\end{table}

\section{Numerical Results\label{sec_expts}}

\subsection{Setup of 2 Experiments}

Proposed algorithm ISPA is evaluated based on three datasets for COVID-19 from \cite{covid19data},
\begin{itemize}
\item \texttt{time\_series\_covid19\_confirmed\_global.csv}
\item \texttt{time\_series\_covid19\_recovered\_global.csv}
\item \texttt{time\_series\_covid19\_deaths\_global.csv}
\end{itemize}
for the time period from January 23 until April 27, 2020. Two experiments were constructed. First, time series for the number of total \emph{net} daily infections were computed by subtracting new daily recoveries from new daily infections. Note that available time series were in some cases reported for multiple different \emph{regions} within the same country. For instance, for Australia it is differentiated between ``Queensland, Australia'', ``New South Wales, Australia'', etc. For consistency in all of the following the term ``country'' will still be maintained. In order to compute \emph{net} daily cases the two aforementioned datasets had to by synchronized by country name, since numbers for \emph{daily new recoveries} were not (yet) always available. After synchronization, the first experiment involved time series for $N=249$ different countries, each of length $T=96$. For the second experiment, the number of daily deaths attributed to COVID-19 was considered. This involved time series for $N=264$ different countries, each of length $T=96$. 

For both experiments the same hyperparameters were used. These are summarized in Table \ref{tab_hyperparam}. In order to emphasize a particular observation, it was intentionally selected $K^\text{max}=N$, i.e., the largest possible number of clusters. It could be observed that for both experiments the number of actual clusters resulting after Step 1 of Algorithm \ref{alg_ISPA} naturally collapsed to $K=14$ and $K=13$ for Experiment 1 and 2, respectively. This indicates that for both cases, net daily infections and daily deaths, available COVID-19 data is highly discriminating.

\subsection{Discussion of Results and Extending Data}

\begin{figure*}
\centering
\input{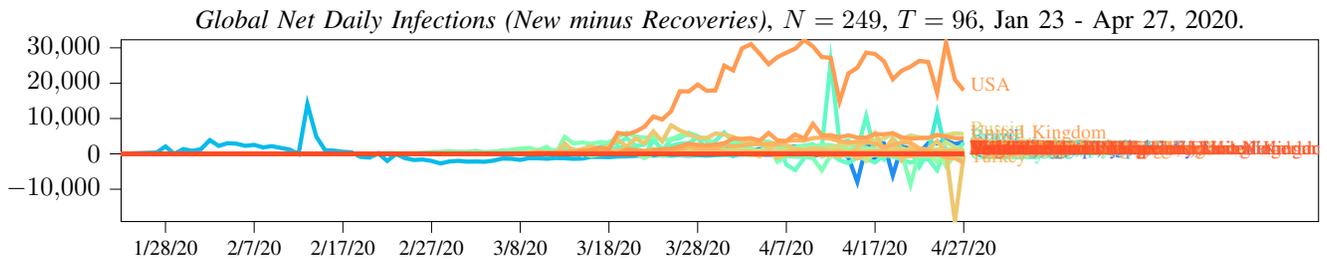}
\caption{Experiment 1. Visualization of the full dataset \cite{covid19data}.}
\label{fig_AllDataCase1}
\end{figure*}

\begin{figure*}
\centering
\input{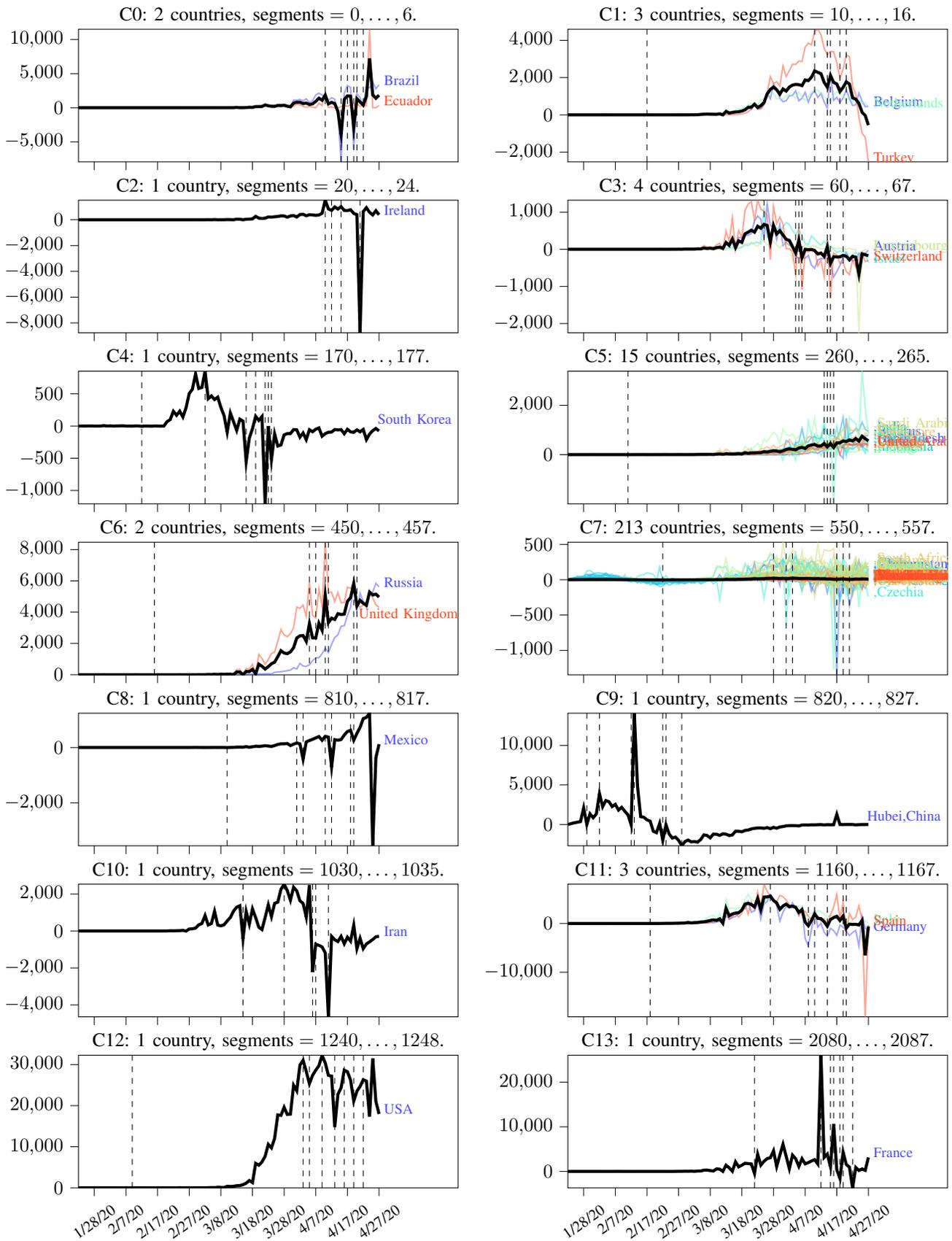}
\caption{Experiment 1. Clustering and segmentation results (shapelet generation according to Steps 1-3 of Algorithm \ref{alg_ISPA}). Note how different and inconsistent shapes are for different countries, with sudden spikes to both the up- and downsides. Segments are separated by black dashed vertical lines. There are 213 countries assigned to cluster C7 with an upper bound of at most 500 net daily infections. The centroid of each cluster is in bold black.}
\label{fig_Case1Clusters}
\end{figure*}

\begin{figure*}
\centering
\input{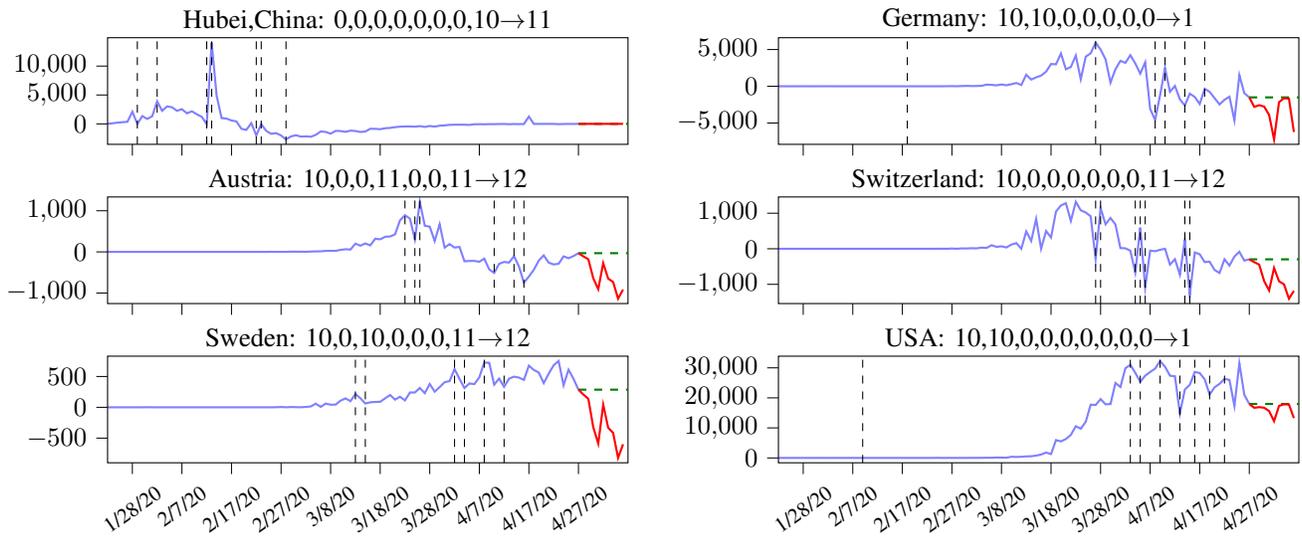}
\caption{Experiment 1 with \emph{net daily infections (new minus recoveries)} on the y-axis for the period from January 23 to April 27 for 6 countries. Labels $\{l_i(p)\}_{i=0}^{P_i}$, followed by $l_i^\text{pred}$ after the arrow, are indicated in the title for each time series $i$. Red and green are 2 prediction trajectories for 10 days ahead, according to Steps 9-15 of Algorithm \ref{alg_ISPA} and according to a constant prediction using the last available measurement, respectively.}
\label{fig_Case1selectedCtries}
\end{figure*}

The results of Experiment 1 are summarized in Fig. \ref{fig_AllDataCase1}, \ref{fig_Case1Clusters} and \ref{fig_Case1selectedCtries}. The following observations are made. 

First, retrieved data \cite{covid19data} is discussed. It is displayed in Fig. \ref{fig_AllDataCase1}. An instant observation is the \emph{global} spread of the disease to 249 countries. COVID-19 can be tested positively also in Timor-Leste, Bhutan and Andorra. The time series for the USA is the most obvious outlier. Another outlier is the blue trajectory with the early spike, which corresponds to Hubei,China. Furthermore, many negative spikes can be detected which indicate more daily recoveries than new infections. This plot globally summarizes 249 different countries. The majority of time series are centered closely around the origin. In this perspective notice the y-axis range. Our planet has a population of 7.8 billion people.  

Second, Fig. \ref{fig_Case1Clusters} summarizes clustering and segmentation results for shapelet generation. The discriminating nature of available data is underlined. Note how, on one hand, 213 countries are assigned to cluster C7 with daily net infections closely fluctuating around the origin. On the other hand, time series for Ireland (C2), South Korea (C4), Mexico (C8), Hubei-China (C9), Iran (C10), USA (C12) and France (C13) each cover single clusters because of their vastly different shapes. Thus, time series greatly differ for different countries. Furthermore, time series are extremely volatile with both sudden positive and negative spikes (see, e.g., France in C13 and Ireland in C2). As a result, because of the volatility, identified shapelets are very short for the time from March onwards, or very long and flat for the time from Januray until March (with the exception of Hubei,China in C9). 

Third, Fig. \ref{fig_Case1selectedCtries} summarizes time series for 6 countries and displays a prediction according to Steps 9-15 of Algorithm \ref{alg_ISPA}. Sprecific countries were selected for the following reasons. Hubei-China was selected since the first confirmed case of COVID-19 has been traced back to 1 December 2019 in its capital Wuhan \cite{cohen2020wuhan}. Germany was selected for its very low death rate (0.01$\%$ deaths per 1 million inhabitants), which will also be discussed later. Austria was selected because of being one of the first European countries to partially have lift closures on April 14. Switzerland was selected because of being, like Austria, an Alpine country, however, in contrast not having eased its lockdown as of April 27. Sweden was selected because of its unique way of handling COVID-19, never imposing a lockdown and keeping schools and restaurants open throughout all times. The USA was selected because of its reporting of the highest numbers of cases. As Fig. \ref{fig_Case1selectedCtries} shows, the trajectories for Germany, Austria and Switzerland are comparable. Since end of March there are on average more daily recoveries than there are new infections. It is very interesting to observe that net daily infections are \emph{plateauing} also for Sweden despite a lack of lockdown. There is \emph{no} exponential growth of reported COVID-19 cases for Sweden. Similarly, net daily infections for the USA are plateauing and maintaining the level as of end of March.


\begin{figure*}
\centering
\input{AllDataCase2.tex}
\caption{Experiment 2. Visualization of the full dataset \cite{covid19data}.}
\label{fig_AllDataCase2}
\end{figure*}

\begin{figure*}
\centering
\input{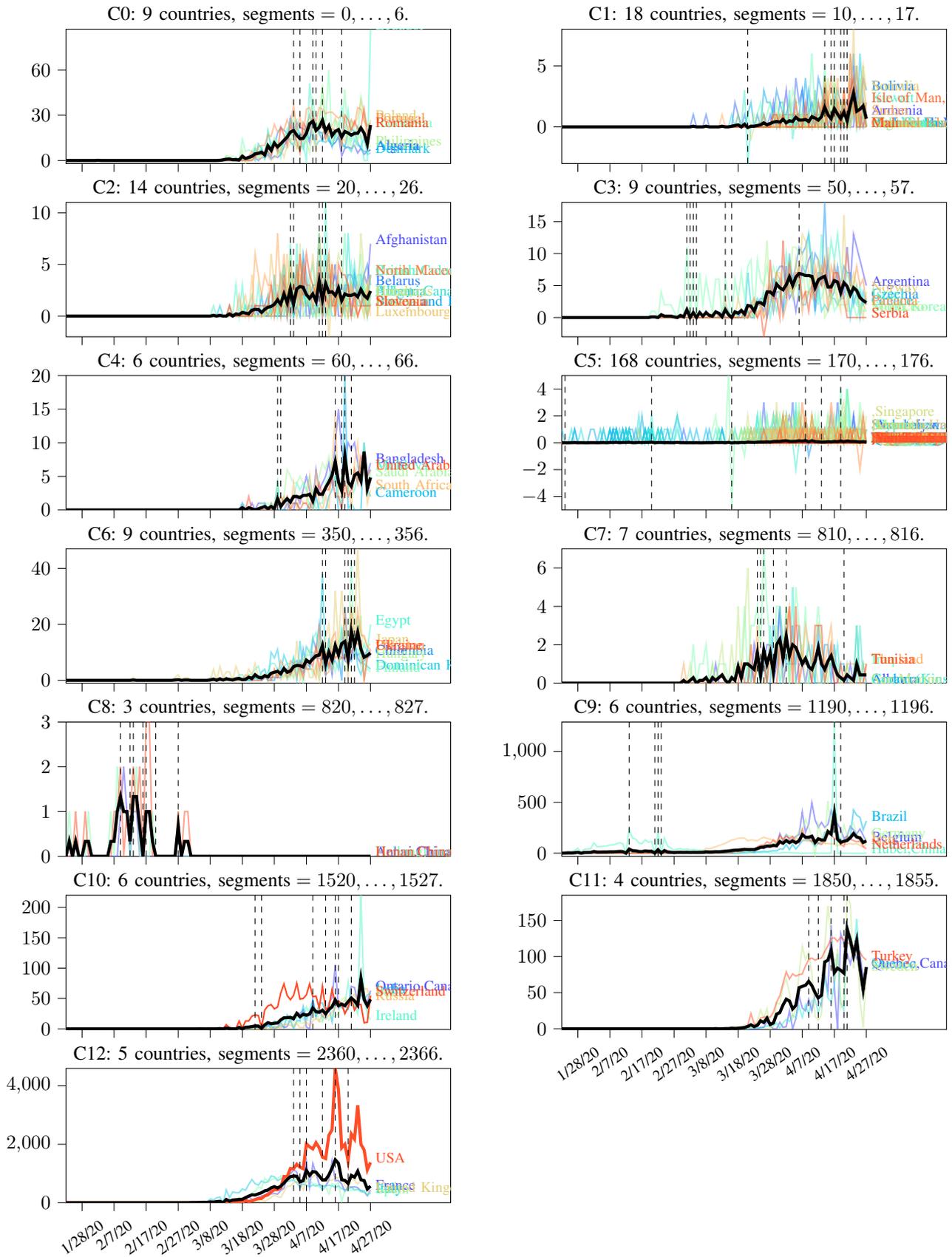}
\caption{Experiment 2. Clustering and segmentation results for \emph{global daily deaths attributed to COVID-19} for the period from January 23 until April 27, 2020. Segments are separated by black dashed vertical lines. Note the different numerical y-axis ranges. The centroid of each cluster is in bold black.}
\label{fig_Case2Clusters}
\end{figure*}

\begin{figure*}
\centering
\input{Case2selectedCtries.tex}
\caption{Experiment 2 with \emph{daily deaths} on the y-axis for the period from January 23 to April 27 for 6 countries. Labels $\{l_i(p)\}_{i=0}^{P_i}$, followed by $l_i^\text{pred}$ after the arrow, are indicated in the title for each time series $i$. Red and green are 2 prediction trajectories for 10 days ahead, according to Steps 9-15 of Algorithm \ref{alg_ISPA} and according to a constant prediction using the last available measurement, respectively.}
\label{fig_Case2selectedCtries}
\end{figure*}

The results of Experiment 2 are summarized in Fig. \ref{fig_AllDataCase2}, \ref{fig_Case2Clusters} and \ref{fig_Case2selectedCtries}. The following observations are made. 

First, retrieved data in Fig. \ref{fig_AllDataCase2} is discussed. Just as for daily net infections, the time series for the USA is the most obvious outlier. The y-axis range is stressed.

Second, Fig. \ref{fig_Case2Clusters} summarizes clustering and segmentation results for shapelet generation. The discriminating nature of available data is underlined. After Step 1 of Algorithm \ref{alg_ISPA} 13 clusters result. The global majority, with 168 countries, is assigned to cluster C5. For this group of countries, the maximum number of daily deaths was \emph{4}. For some countries negative numbers are reported (data is used as downloaded without correction). The vast differences in y-axis ranges are stressed. There are 5 clusters (C1, C2, C5, C7, and C8) with single-digit daily deaths for a total of 210 countries. There are 4 clusters (C0, C3, C4 and C6) with at most double-digit daily deaths for a total of 33 countries. Because of the volatility in recorded data, identified shapelets are either very short for the time from March onwards, or very long and flat for the time from January until March. 

Third, Fig. \ref{fig_Case1selectedCtries} summarizes time series for 6 countries and displays a prediction according to Steps 9-15 of Algorithm \ref{alg_ISPA}. The trajectories for Germany, Austria, Switzerland and the USA are comparable by shape, but differ by y-axis ranges. For Austria (with a population of 8.9 millions) the \emph{maximum} number of daily deaths attributed to COVID-19 for the the time period from January until April 27 was \emph{30}. For the USA (with a population of 328.2 millions) the corresponding number was above 4000. Sweden has a highly fluctuating number with an upper bound below 200.

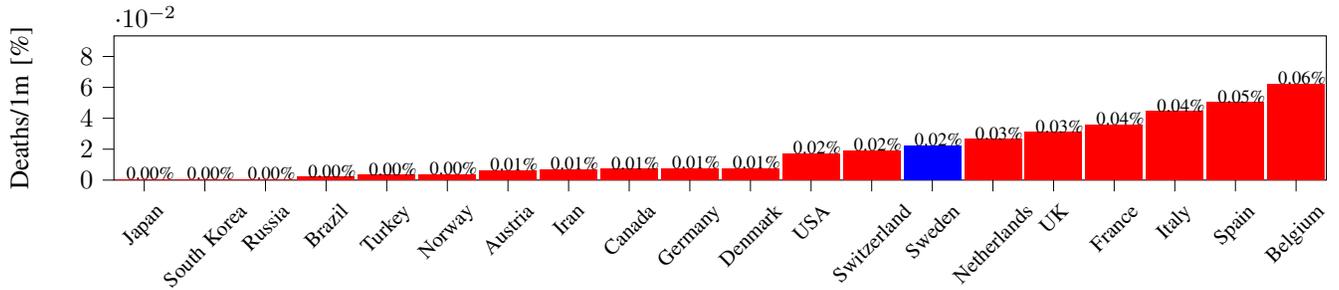
\begin{figure*}
\centering
\begin{tikzpicture}

\begin{axis}[
tick align=outside,
tick pos=left,
width=17.7cm,
height=3.5cm,
x grid style={white!69.0196078431373!black},
xmin=0, xmax=20,
xtick style={color=black},
xtick={0.5,1.5,2.5,3.5,4.5,5.5,6.5,7.5,8.5,9.5,10.5,11.5,12.5,13.5,14.5,15.5,16.5,17.5,18.5,19.5},
xticklabel style = {rotate=45.0,scale=0.8},
xticklabels={Japan,South Korea,Russia,Brazil,Turkey,Norway,Austria,Iran,Canada,
Germany,Denmark,USA,Switzerland,Sweden,Netherlands,UK,France,Italy,Spain,Belgium},
y grid style={white!69.0196078431373!black},
ylabel={Deaths/1m [\%]},
ylabel style = {scale=1},
ymin=0, ymax=0.0933,
ytick style={color=black},
]
\draw[draw=none,fill=red] (axis cs:0.025,0) rectangle (axis cs:0.975,0.0003);
\draw[draw=none,fill=red] (axis cs:1.025,0) rectangle (axis cs:1.975,0.0005);
\draw[draw=none,fill=red] (axis cs:2.025,0) rectangle (axis cs:2.975,0.0005);
\draw[draw=none,fill=red] (axis cs:3.025,0) rectangle (axis cs:3.975,0.0021);
\draw[draw=none,fill=red] (axis cs:4.025,0) rectangle (axis cs:4.975,0.0034);
\draw[draw=none,fill=red] (axis cs:5.025,0) rectangle (axis cs:5.975,0.0038);
\draw[draw=none,fill=red] (axis cs:6.025,0) rectangle (axis cs:6.975,0.0061);
\draw[draw=none,fill=red] (axis cs:7.025,0) rectangle (axis cs:7.975,0.0069);
\draw[draw=none,fill=red] (axis cs:8.025,0) rectangle (axis cs:8.975,0.0072);
\draw[draw=none,fill=red] (axis cs:9.025,0) rectangle (axis cs:9.975,0.0073);
\draw[draw=none,fill=red] (axis cs:10.025,0) rectangle (axis cs:10.975,0.0074);
\draw[draw=none,fill=red] (axis cs:11.025,0) rectangle (axis cs:11.975,0.0172);
\draw[draw=none,fill=red] (axis cs:12.025,0) rectangle (axis cs:12.975,0.0192);
\draw[draw=none,fill=blue] (axis cs:13.025,0) rectangle (axis cs:13.975,0.0225);
\draw[draw=none,fill=red] (axis cs:14.025,0) rectangle (axis cs:14.975,0.0264);
\draw[draw=none,fill=red] (axis cs:15.025,0) rectangle (axis cs:15.975,0.0311);
\draw[draw=none,fill=red] (axis cs:16.025,0) rectangle (axis cs:16.975,0.0357);
\draw[draw=none,fill=red] (axis cs:17.025,0) rectangle (axis cs:17.975,0.0446);
\draw[draw=none,fill=red] (axis cs:18.025,0) rectangle (axis cs:18.975,0.0503);
\draw[draw=none,fill=red] (axis cs:19.025,0) rectangle (axis cs:19.975,0.0622);
\draw (axis cs:0.1,0.0003) node[
  scale=0.7,
  anchor=base west,
  text=black,
  rotate=0.0
]{0.00\%};
\draw (axis cs:1.1,0.0005) node[
  scale=0.7,
  anchor=base west,
  text=black,
  rotate=0.0
]{0.00\%};
\draw (axis cs:2.1,0.0005) node[
  scale=0.7,
  anchor=base west,
  text=black,
  rotate=0.0
]{0.00\%};
\draw (axis cs:3.1,0.0021) node[
  scale=0.7,
  anchor=base west,
  text=black,
  rotate=0.0
]{0.00\%};
\draw (axis cs:4.1,0.0034) node[
  scale=0.7,
  anchor=base west,
  text=black,
  rotate=0.0
]{0.00\%};
\draw (axis cs:5.1,0.0038) node[
  scale=0.7,
  anchor=base west,
  text=black,
  rotate=0.0
]{0.00\%};
\draw (axis cs:6.1,0.0061) node[
  scale=0.7,
  anchor=base west,
  text=black,
  rotate=0.0
]{0.01\%};
\draw (axis cs:7.1,0.0069) node[
  scale=0.7,
  anchor=base west,
  text=black,
  rotate=0.0
]{0.01\%};
\draw (axis cs:8.1,0.0072) node[
  scale=0.7,
  anchor=base west,
  text=black,
  rotate=0.0
]{0.01\%};
\draw (axis cs:9.1,0.0073) node[
  scale=0.7,
  anchor=base west,
  text=black,
  rotate=0.0
]{0.01\%};
\draw (axis cs:10.1,0.0074) node[
  scale=0.7,
  anchor=base west,
  text=black,
  rotate=0.0
]{0.01\%};
\draw (axis cs:11.1,0.0172) node[
  scale=0.7,
  anchor=base west,
  text=black,
  rotate=0.0
]{0.02\%};
\draw (axis cs:12.1,0.0192) node[
  scale=0.7,
  anchor=base west,
  text=black,
  rotate=0.0
]{0.02\%};
\draw (axis cs:13.1,0.0225) node[
  scale=0.7,
  anchor=base west,
  text=black,
  rotate=0.0
]{0.02\%};
\draw (axis cs:14.1,0.0264) node[
  scale=0.7,
  anchor=base west,
  text=black,
  rotate=0.0
]{0.03\%};
\draw (axis cs:15.1,0.0311) node[
  scale=0.7,
  anchor=base west,
  text=black,
  rotate=0.0
]{0.03\%};
\draw (axis cs:16.1,0.0357) node[
  scale=0.7,
  anchor=base west,
  text=black,
  rotate=0.0
]{0.04\%};
\draw (axis cs:17.1,0.0446) node[
  scale=0.7,
  anchor=base west,
  text=black,
  rotate=0.0
]{0.04\%};
\draw (axis cs:18.1,0.0503) node[
  scale=0.7,
  anchor=base west,
  text=black,
  rotate=0.0
]{0.05\%};
\draw (axis cs:19.1,0.0622) node[
  scale=0.7,
  anchor=base west,
  text=black,
  rotate=0.0
]{0.06\%};
\end{axis}

\end{tikzpicture}
\caption{Overview of 20 countries ranked by \emph{deaths} (attributed to COVID-19) \emph{per 1 million inhabitants} as of April 28, 2020 \cite{worldometers}. Sweden is emphasized because of its different approach of handling the pandemic in contrast to other countries with avoidance of any lockdown.}
\label{figBarDeathsPer1m}
\end{figure*}
\begin{figure*}
\centering
\input{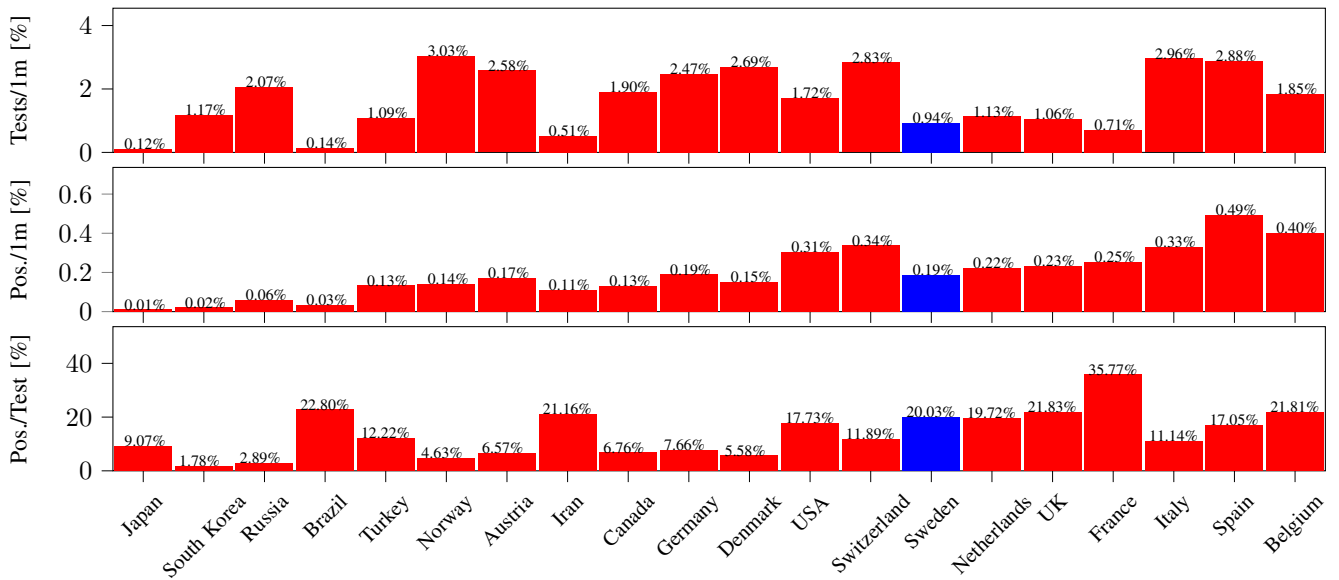}
\caption{Summary of 3 different ratios for 20 countries ranked according to Fig. \ref{figBarDeathsPer1m} as of April 28, 2020. ``Pos.'' is abbreviated for ``Positive''. Note that reported numbers are \emph{accumulated over the entire duration} of the pandemic since its start\cite{worldometers}.}
\label{fig_barSubplots}
\end{figure*}

Because of the variability, volatility and differences of time series for countries, extending statistics are provided in Fig \ref{figBarDeathsPer1m} and \ref{fig_barSubplots}. The following observations are made.

First, it is astonishing to see Sweden ranked only 7th, approximately on par with Switzerland and USA, for \emph{deaths per 1 million inhabitants} as shown in Fig. \ref{figBarDeathsPer1m}. This is highly surprising given its approach of handling COVID-19 \emph{without} lockdown in contrast to all higher ranked countries (Netherlands, UK, France, Italy, Spain and Belgium). Remarkably, the normalized death rate for Belgium is three times higher than that of Sweden. 

Second, equally surprising is the number of \emph{positive tests per 1 million inhabitants} for Sweden with $0.19\%$, which is even significantly lower than the $0.31\%$ for the USA. 

Third, there is a large variability in the number of \emph{positive tests per total number of tests}. For example, for a COVID-19 test in France there is a 35.77$\%$ likelihood that the test is positive. In neighboring Italy the likelihood is only 11.14$\%$.


A brief comment about the \emph{basic reproduction number} $R_0$ is made, which recently gained significant attention in the media and is used by goverments as gauge to measure effectiveness of their restrictions \cite{nytR0}. $R_0$ represents the average number of people that is infected from an infected person. For COVID-19, $R_0$ is estimated to be between 2 and 2.5, and the serial interval (average time between each successive infection) is estimated to lie between 4 to 4.5 days. Assuming average values for $R_0$ and the serial interval, the total \emph{accumulated} number of human beings, $H(T)$, infected up until time $T$ can be calculated as $
H(T) = \sum_{t=0}^{T} R_0^{\frac{t}{4.25}}$, with $R_0=2.25$. For the time period from January 23 until April 29 this corresonds to $T=98$ and $H(T)=628.4$m. However, according to \cite{worldometers} the accumulated number of human beings that came in contact since outbreak of COVID-19 is only $3.2$m as of April 29. Hence, $H(T)$ is larger by a factor of \emph{195}. To develop this further, the first confirmed case of COVID-19 has been traced back to December 1, 2019 in Wuhan \cite{cohen2020wuhan}. As of April 29, this would correspond to $T=152$, and thus $H(T)=18750.7$ billion, which is more than \emph{twice} the total world population. Obviously, above model is a simplification, that does, e.g., not account for damping effects such as immunization etc. On the other hand, France (and many other western countries) announced strict confinement measures only on March 17, which would imply that COVID-19 could have spread unchecked for almost three months. India, with a population of 1.3 billion, ordered a nationwide lockdown only on March 24.

While $R_0$ is related to the number of \emph{infections}, a brief comment about available data for \emph{deaths attributed to COVID-19} is made. There are many examples such as Morton's. ``Morton's death certificate said Morton's cause of death was COVID-19'' \cite{109years}. Morton was 109 years old.

Finally, while there are many uncertainties regarding available COVID-19 data, this is clearly not the case for measurements of the economic ramifications due to governmental restrictions and lockdowns. Here, numbers are well measurable. In the USA alone, jobless claims total more than 30 million after week 6 of lockdown \cite{bloomberg30m}. Similar devastations in other countries are omitted for brevity. The global cost of COVID-19 may reach 4.1 Trillion \cite{trillions}. This excludes social cost. Furthermore, the ``fear of COVID-19'' may be just as deadly as the disease itself. It is entirely plausible that more people die (in particular also the elderly) because of social isolation or eoconomic hardship rather than from COVID-19.  In this perscpective, consider \cite{nytHeart}: ``Where are all the patients with heart attacks and stroke? They are missing from our hospitals.''

\section{Conclusion\label{sec_concl}}

This paper presented an Integrated Summarization and Prediction Algorithm (ISPA) for the handling of a set of multiple related time series. It was applied to real-world COVID-19 data in two experiments; First, for global net daily infections (new daily infections minus daily recoveries), and, second, for global daily deaths for the period from January 23 until April 27, 2020. This resulted in datasets for 249 and 264 countries, respectively. While it was found that predictions provided by ISPA were not very useful due to very volatile and short COVID-19 data, a variety of insights drawn from the summarization of global data were discussed. Analysis on extending data, including normalized ratios adjusted for population sizes, was also presented.

If available data is to be believed, the worst in terms of number of daily net infections and daily deaths attributed to COVID-19 seems to have past in the majority of countries as of April 27, 2020. However, various anomalies detected make believe that reported data is likely not accurate.

There is a strong discrepancy between, on the one hand, \emph{fully measurable and known} economically and socially devastating impacts of extended lockdowns, and, on the other hand, a lot of \emph{inexplicabilities and controversies} about COVID-19 and its available data. To name just one example, why do Belgium and France have \emph{3} and \emph{2} times more \emph{deaths per 1 million inhabitants} in comparison to Sweden, eventhough Belgium and France both entered very strict lockdowns on March 18, where ``people were allowed to go out in an emergency or to a supermarket, pharmacies and doctors, but any gatherings were prohibited''? If country-wide lockdowns helped so much, should Sweden not be off \emph{much} worse since it never enforced a lockdown? 

In view of the trade-off between protective restrictions, an urge to reopen for the saving of livelihoods, and in view of the unreliability of official data, it is believed that a possible health-first and justifiable solution is a decentralized approach in which heads or consortia of doctors and nurses of local hospitals decisively guide local reopening debates. It is believed that this group of professionals is currently largely and undeservedly underrepresented in many public discussions (that focus on predictions of virologists instead), eventhough these bear the brunt and offer the more realistic, real-time and localized state estimate on the disease.

%
\nocite{*}
\bibliographystyle{ieeetr}
\bibliography{myref}
%
\end{document}